\documentclass{aip-cp}

\usepackage[numbers]{natbib}
\usepackage{rotating}
\usepackage{graphicx}
\usepackage{mathrsfs}
\usepackage{amssymb}
\usepackage{graphicx}
\usepackage[normalem]{ulem}
\usepackage[dvips]{color}
\usepackage{bm}
\usepackage{longtable}
\usepackage{times}

\renewcommand\sout{\bgroup \color{red} \ULdepth=-.5ex \ULset}

\renewcommand{\rm}[1]{\textrm{#1}}

\begin{document}

\title{Trinity of Strangeon Matter}
\author[aff1,aff2]{Renxin Xu}
\affil[aff1]{School of Physics and Kavli Institute for Astronomy and Astrophysics, Peking University, Beijing 100871, China,}
\affil[aff2]{State Key Laboratory of Nuclear Physics and Technology, Peking University, Beijing 100871, China; r.x.xu@pku.edu.cn}
\maketitle

\begin{abstract}
Strangeon is proposed to be the constituent of bulk strong
matter, as an analogy of nucleon for an atomic nucleus.
The nature of both nucleon matter (2 quark flavors, {\it u} and {\it
d}) and strangeon matter (3 flavors, {\it u},  {\it d} and {\it s})
is controlled by the strong-force, but the baryon number of the
former is much smaller than that of the latter, to be separated by a
critical number of $A_{\rm c}\sim 10^9$.
While micro nucleon matter (i.e., nuclei) is focused by nuclear
physicists, astrophysical/macro strangeon matter could be
manifested in the form of compact stars (strangeon star), cosmic
rays (strangeon cosmic ray), and even dark matter (strangeon dark
matter).
This trinity of strangeon matter is explained, that may impact
dramatically on today's physics.
\end{abstract}

\vspace{0.6cm}
{\em Symmetry does matter: from Plato to flavour}.
Understanding the world's structure, either micro or macro/cosmic, is
certainly essential for Human beings to avoid superstitious belief as well as to
move towards civilization.
The basic unit of normal matter was speculated even in the
pre-Socratic period of the Ancient era (the basic stuff was
hypothesized to be indestructible ``atoms'' by Democritus), but it
was a belief that symmetry, which is well-defined in mathematics,
should play a key role in understanding the material structure, such
as the Platonic solids (i.e., the five regular convex polyhedrons).
In this contribution, we are addressing that quark flavour-symmetry restoration from {\bf 2} ({\it u} and {\it d} quarks) to {\bf 3} ({\it u}, {\it d} and {\it s} quarks) should be essential in making compressed baryonic matter when normal {\bf 2}-flavoured matter inside an evolved massive star is squeezed during a core- collapse supernova.

Nature loves symmetry, but with symmetry breaking at negligible scale, $\delta$.
For an example related to the topics of this conference, stable
nuclei (i.e., {\it nucleon} matter) are symmetric with {\bf
2}-flavors of quarks (i.e., isospin symmetry, or namely nuclear
symmetry energy), while the symmetry is broken at a level of $\delta
< 0.2$ (for the most heavy stable nucleus, $^{208}$Pb, $\delta =
(126-82)/208 = 0.2$; for the most binding nucleus, $^{56}$Fe,
$\delta = (30-26)/568 = 0.07$).
For one ``gigantic'' nucleus where a huge number of normal {\bf
2}-flavoured nuclei merge, it is proposed that {\bf 3}-flavoured
{\it strangeon} (former name: quark cluster~\cite{Xu03}), rather
than {\bf 2}-flavoured nucleon, might serve as the building block.
For bulk/macroscopic strong matter, the {\bf 2}-flavour symmetry of
nucleon matter has to be broken significantly (thus so-called
symmetry energy contributes a lot) due to neutronization, but {\bf 3}-flavour symmetric
strangeon matter could only be broken negligibly after strangeonization.
Therefore, bulk strong matter would be strangeon matter if Nature really likes symmetry (i.e., a principle of flavor maximization~\cite{Xu18}).

In 1932, an idea of gigantic nucleus was tried by Lev
Landau~\cite{Landau32}, which develops then, especially after the
discovery of radio pulsars, to be very elaborated models of
conventional neutron stars (i.e, nucleon stars, with asymmetric
flavours of quarks and complex inner structures, in the mainstream).
Alternatively, {\bf 3}-flavour symmetry could be restored when
normal matter density becomes so great that {\bf 2}-flavoured nuclei
come in close contact, forming bulk strong matter.
This alternative solution to the equation of state (EoS) of dense
matter at supranuclear density is truly in the regime of ``old
physics'' (i.e., not beyond the standard model of particle physics),
but may have particular consequences for us to understand
additionally dark matter and even ultra-high energy cosmic rays.
The Occam's razor nature of strangeon matter may also mirror the
Chinese Taoist philosophy: Da Dao Zhi Jian (the greatest truths are
the simplest, or, simplicity is always universality).
In this spirit, I wish the strangeon conjecture is too simple to be
ruled out in the future.

\section{What is a strangeon?}\label{strangeon}

Strange quark, $s$, is actually {\em not} strange, it is named after
a quantum number of strangeness (conserved in strong interaction but
changeable in weak interaction) discovered via cosmic ray
experiments in 1947.
There are totally six flavours of quarks in the standard model of
particle physics, half ($u$, $d$ and $s$, with current masses
smaller than $\sim 0.1$ GeV) are light and the other half ($c$, $t$
and $b$, masses $m_{\rm{heavy}}>1$ GeV) are heavy.
It is worth noting that only two flavours of valence quarks, $u$ and
$d$, are responsible to the common material of the world today, and
then we should not be surprised that the third light flavour was
initially named as strange quark, $s$.
We usually neglect heavy flavours of quarks in the study of dense matter at a few or around nuclear density, only because the energy scale there, estimated with the Heidelberg's relation, would be order of $E_{\rm{scale}}\sim \hbar c/\Delta x\sim 0.5$ GeV $<m_{\rm{heavy}}$, where the separation between quarks is $\Delta x\sim 0.5$ fm.
It is shown that this energy is higher than the mass difference between strange and up/down quarks ($\Delta m_{\rm{uds}}\sim 0.1$ GeV), $E_{\rm{scale}}\gg \Delta m_{\rm{uds}}$, and we may expect a {\bf 3}-flavoured universe. But why is our world {\bf 2}-flavoured? This is a topic related to the nuclear symmetry energy, focused in this meeting.

Microscopic strong matter (i.e., normal nuclei in atoms) should be
{\bf 2}-flavoured, even Nature may love a principle of quark flavour
maximization, because of a {\em neutrality problem}.
It is evident that strong matter with 2-flavour symmetry is positively charged (this is the reason that an atomic nucleus is electrically positive), and electrons (mass $m_{\rm e}\simeq 0.5$ MeV) would have to emerge with a number about half the baryon number.
Nevertheless, these electrons do  not matter critically for a
micro-nuclus because of smallness (i.e., normal nuclei are too small
in size), so that all the electrons are outside and
non-relativistic.
Therefore, for normal matter, {\bf 3}-flavoured nuclei would be
energetically unstable due to weakly conversing $s$- to $u/d$-quarks
for $m_{\rm s}-m_{\rm{ud}}\gg m_\rm{e}$, with $m_{\rm{ud}}$ the mass
of either $u$- or $d$-quarks, but {\bf 2}-flavour symmetry keeps
even though $d$-quark is more massive than $u$-quark (isospin
symmetry) as $E_{\rm{scale}}\gg \Delta m_{\rm{ud}}$ (note: $m_{\rm
u}$ = 2.15 MeV and $m_{\rm d}$ = 4.70 MeV, $\Delta
m_{\rm{ud}}=m_{\rm d}-m_{\rm u}$, while $m_{\rm s}$ = 93.8 MeV,
determined from lattice gauge simulations).
This hints that Nature may love flavour maximization.

Macroscopic strong matter (i.e., a huge number of normal nuclei
merge to form a gigantic nucleus), however, should be {\bf
3}-flavoured if Nature really love the flavour maximization
principle, without the neutrality problem if {\bf 3}-flavour
symmetry is restored.
Strangeness has already been included to understand the nature of strong matter since 1970s, but particular attention has been paid for the case of free quarks~\cite{Bodmer1971,Witten1984} (so-called strange {\em quark} matter).
Nevertheless, the perturbative quantum chromo-dynamics (QCD), based
on asymptotic freedom, works well only at energy scale of
$\Lambda_\chi>1$ GeV, and then the state of pressure-free strong
matter should be relevant to non-perturbative QCD because of
$E_{\rm{scale}}<\Lambda_\chi$, exactly a similar case of normal
atomic nuclei.
A conjecture of ``condensation'' in position space (constituent unit: strange quark cluster~\cite{Xu03}), rather than in momentum space for a color super-conducting state, was thus made for cold  matter at supra-nuclear density. The strange cluster is renamed {\em strangeon}, being coined by combining ``strange nucleon'' for the sake of simplicity~\cite{xg2017,Wang2017}, as illustrated in Fig.~\ref{strangeon}.
\begin{figure}[h!]
\centering
  \includegraphics[width=12cm]{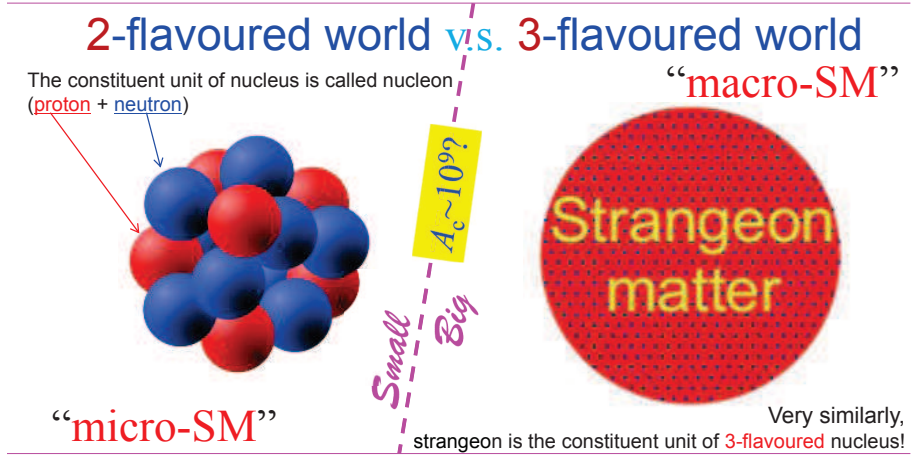}
  \caption{While normal atom matter is controlled by electromagnetic interaction
  (we may simply call by a name of {\em electric} matter), the nature of bulk {\em strong}
  matter (e.g., inside pulsar-like compact star) is determined by the fundamental
  strong force. It is conjectured that microscopic strong matter (micro-SM) is
  {\bf 2}-flavoured (basic unit: nucleon) but macroscopic strong matter (macro-SM)
  is {\bf 3}-flavoured (basic unit: strangeon), with a critical baryon number of
  $A_{\rm c}$. One could estimate $A_{\rm c}\simeq \lambda_{\rm c}^3/\rm{fm}^3\sim 10^9$,
  where the electron Compton wavelength $\lambda_{\rm c} = h/(m_{\rm e}c)=2.4\times 10^3$ fm.}
  \label{strangeon}
\end{figure}
The color coupling of strong matter at zero pressure, both microscopic and macroscopic, should be so strong that quarks are localized (in nucleon and strangeon, respectively).
Nonetheless, the quantum effect of strangeon would be weaker than that of nucleon because the former is more massive than the latter, and it is proposed that cold strangeon matter should be in a solid state if the kinematic thermal energy is much lower than the interaction energy between strangeons~\cite{Xu03}.

Electron, as fundamental lepton that does not undergo the strong interaction
(flavour unchangeable) but does participate in the weak one (flavour
changeable), may play a key role in the determination
of the critical baryon number ($A_{\rm c}$) to differentiate macro-
from micro-strong matter.
This is, in fact, not a new concept, but was initiated by Lev
Landau~\cite{Landau32}, who speculated that a doublet (called neutron later) could
form via combining closely a proton and an electron because ``{\em
we have always protons and electrons in atomic nuclei very close
together}'' and thought that ``{\em the laws of quantum mechanics
(and therefore of quantum statistics) are violated}'' in those ``{\em pathological regions}'', before the discovery of neutron and the recognition of the weak interaction.
Landau then expected ``that this must occur when the density of
matter becomes so great that atomic nuclei come in close contact,
forming one gigantic nucleus''~\cite{Landau32}.
In the word of modern physics, Landau provided a way of {\em
neutronization}, $e^-+p\rightarrow n+\nu_{\rm e}$, to kill
energetic electrons during squeezing normal baryonic matter.
Nevertheless, there is another way, so-called {\em
strangeonization}, to eliminate those relativistic electrons in the
standard model of particle physics~\cite{xg2017}, a way to be {\bf
3}-flavor-symmetric (note: it is extremely {\bf 2}-flavor-asymmetric
after neutronization). One can then judge if Nature loves flavor
maximization.
It is worth noting that both neutronization and strangeonization are
of the weak interaction that could work for changing quark flavour.
Certainly, we have to have a reliable way to quantitatively
evaluate the truth of either neutronization or strangeonization, and we
are trying an effort to construct a linked-bag model for condensed strong-matter, to be announced through other publications in the future.
In view of the importance of electron's role, therefore, we may use
the electron Compton wavelength, $\lambda_{\rm c} = h/(m_{\rm e}c)$,
to mark the boundary between micro- and macro- strong matter, and
the critical baryon number could thus be $A_{\rm c}\simeq
\lambda_{\rm c}^3/\rm{fm}^3\sim 10^9$, as annotated in Ref.~\cite{Xu2015}.

\subsection{Trinity of Strangeon Matter}\label{trinity}

For atom/molecular matter, compared to gases where the
in-between interactions are negligible, condensed matter is of liquid or solid
where the interaction is strong enough to make atomic units
cohesive.
Although we have a lot problems in understanding liquid (``the liquid state, in some ways, has no right to exist ''~\cite{Tabor91}), it is find
that the interaction potential, $V(r)$, between the building units with
separation $r$ in normal condensed matter can be represented as the sum of
long-distant attraction and short-distant repulsive,
in a resultant form of Mie's potential,
\begin{equation}
V(r) = -{A\over r^m}+{B\over r^n},
\end{equation}
where $m<n$ (both are positive integers), and $A$ and $B$ are
positive constants. Specifically, Lennard-Jones' potential is practically employed in modelling the in-between interaction (e.g., van der Waals' interactions), with $m=6$ and $n=12$,
\begin{equation}
V(r) = 4\epsilon [-({\sigma\over r})^6+({\sigma\over r})^{12}],
\label{6-12}
\end{equation}
where two parameters, $\epsilon$ and $\sigma$, characterize the Lennard-Jones 6-12 potential, the former is associated with the interaction energy and the latter measures the separation between units.
It is evident that the potential curve crosses the $r$-axis at $r=\sigma$, but the potential well minimizes at $r=2^{1/6}\sigma\simeq 1.12\sigma$, with a depth of $-\epsilon$.

For nucleon/strangeon matter, the Lennard-Jones model would apply
too, in which the interaction between the strong units (nucleon or strangeon) is also found to be
similar to that between atoms/molecules, because the strong units are colorless, as in the case of chargeless atoms.
Experimentally, for nucleons, while the attraction force represents
the so-called nuclear force, the existence of a hard core (i.e., the
repulsion at short distance) is opaque to theoretical analysis which
is essentially a non-perturbative consequence and would be certainly
crucial to the nuclear physics. The hard core is entirely empirical,
but the strong internucleon forces (including
the hard-core)  could be reproduced via numerical lattice QCD~\cite{Wilczeck2007}.
Then it would not be surprising that both nucleon and atom could
share a common nature of 6-12 potential.
For strangeon matter, we may expect also Lennard-Jones-like
interstrangeon force, so that condensed strangeon matter could exist
in nature, with baryon number $A>A_{\rm c}\sim 10^9$.
It is found, ten years ago, that the EoS with interactions of
Eq.~\ref{6-12} is very stiff~\cite{LX09}, before the discovery of
massive pulsars around $2M_\odot$.

In analogy to condensed electronic matter, condensed strong matter
could have homogenous density for the case of gravity free in which
the gravitational energy is much smaller than it's binding energy,
but can also have density gradient for the case of stellar objects.
Obviously, this implies that the mass-radius ($M-R$) relation
changes from $M\propto R^3$ (i.e., $M/R^3=$ const.) at low-mass to
higher values of $M/R^3$ at high-mass.
Although in ``old'' physics, strangeon matter does matter, with
particular consequences in observations, and we could expect with
confidence that condensed strangeon matter would be manifested in
various different forms, as long as $A>A_{\rm c}\simeq 10^9$, such as strangeon
cosmic ray ($A\sim 10^{10}$), strangeon dark matter ($A\sim
10^{30}$), strangeon planet ($A\sim 10^{54}$) and strangeon star
($A> \sim 10^{57}$), all of which would have dramatic consequences
in today's physics.

\vspace{2mm}
\noindent%
1. {\em Strangeon Star/Planet}.
Strangeon and strangeon star have already been introduced
briefly~\cite{LaiXu17,LuXu17}, with significant attention paid to
the peculiar observational features related to both the surface
condition and the global structure of strangeon stars.
Certainly it is worthwhile to identify a strangeon star by advanced
facilities, but much work is needed in order to take advantage of
the unique opportunities those facilities will provide.
In addition to the astrophysical features previously discussed, we
will note here three points as indicated in the following.

\begin{figure}[h!]
\centering
  \includegraphics[width=8cm]{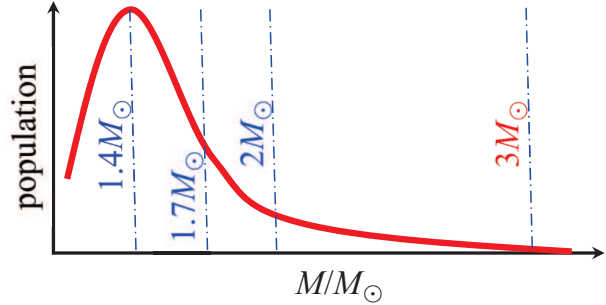}
  \caption{A conjectured population as a function of initial mass for strangeon stars created after core-collapse supernova.
  It is suggested to detect pulsar's masses between $2M_\odot$ and $3M_\odot$, especially with the most sensitive FAST (Five-hundred-meter Aperture Spherical
  radio Telescope). This discovery may provide strong evidence for a very stiff EoS of supranuclear matter.}
  \label{mass}
\end{figure}

First, because of two reasons (I: massive strangeons to be
non-relativistic, II: interstrangeon hard core illustrated in
Eq.~\ref{6-12}), the EoS of strangeon matter is so stiff that the
mass limit, $M_\rm{max}$, could reach and even be hight than
$3M_\odot$.
It is conventionally thought that the pulsar mass spectrum peaks at $\sim 1.4M_\odot$~\cite{tc99}, but there is evidence for neutron star born massive~\cite{Tauris11} (the initial mass could be $>1.7M_\odot$ for PSR J1614-2230, the first massive pulsar detected).
Although it is worthwhile to search pulsars with higher masses
(e.g., \cite{0740}), strangeon star's birth masses after
core-collapse supernova could be far smaller than the limit $\sim
3M_\odot$.
It is then speculated that core-collapse supernova would produce strangeon stars with mass around $1.4M_\odot$, but the population of strangeon star decreases as the mass increases, as illustrated in Fig.~\ref{mass}.
Therefore, we can uncommonly discover a nascent strangeon star to be
massive ($M>M_\rm{max}$) enough to collapse quickly into a black
hole after a supernova, and a relevant study of the population
synthesis should be welcome.
Nevertheless, binary compact star merger and ultraluminous X-ray source provide two fantastic channels to create very massive strangeon object ($M\lesssim M_\rm{max}$), with multi-messenger astronomy.
It is also worth noting that, in contrast to objects with high
masses, hunting a low-mass strangeon star ($0.01\sim 0.5)M_\odot$)
or a strangeon planet ($<10^{-2}M_\odot$) is also crucial for their
identification, as discussed in Refs.~\cite{XW03,Horvath12,Xu14}.

Second, though strangeon star model survives the scrutiny of GW 170817~\cite{Lai18,LZX19} (the maximum mass would be $M_\rm{max}\sim 3M_\odot$, but the tidal deformability of a $1.4M_\odot$ star could be as low as $\Lambda\sim 200$), it is
still a matter of debate whether it could pass the test of kilo-nova
(KN) observations of merging compact objects.
In the regime of neutron star merger, the compact star is supposed
to be made almost entirely of neutrons, and merging binary stars could produce neutron-rich ejecta in which r-process nucleosynthesis will happen.
From the flavour-symmetric point of view, this neutron kilo-nova
(NKN) scenario is for changing {\bf 2}-flavoured asymmetry to almost
{\bf 2}-flavoured symmetry, by the reverse mode of neutronization
(or simply called inverse-neutronization).
\begin{figure}[h!]
\centering
  \includegraphics[width=12cm]{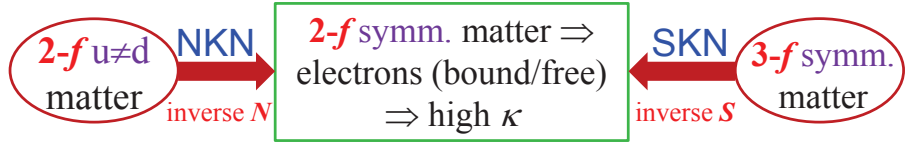}
  \caption{%
  Neutron star and strangeon star are two concepts, with which we can understand the astrophysical manifestations of pulsar-like compact stars. Compact star merger provides a fantastic way to test
  neutron and strangeon star models. Besides the dynamical tidal
  deformability, $\Lambda$, kilonova (KN) observations are valuable
  for us to test models, though being model-dependent.
  While the neutron kilo-nova (NKN) is of symmetry restoration for two flavours, the
  strangeon kilo-nova (SKN) is to change the flavour-symmetric number (from {\bf 3}
  to {\bf 2}). Both KN scenarios can release considerable amount of electrons, bound
  or free, eventually resulting in an environment with high opacity, $\kappa$.
  }%
  \label{KN}
\end{figure}
In the regime of strangeon star merger, however, the strangeon
kilo-nova (SKN) scenario is for changing {\bf 3}-flavoured symmetry
to almost {\bf 2}-flavoured symmetry, by the reverse mode of
strangeonization (or simply called inverse-strangeonization), with
regard to light strangeon nuggets ejected ($A<A_{\rm c}$), as
summarized in Fig.~\ref{KN}.
Part of strangeon nuggets with $A>A_{\rm c}$ will fly from
the KN site to, maybe, the Earth's atmosphere, and an air-shower of
cosmic ray occurs.
Strangeon nuggets with $A<A_{\rm c}$ decay quickly via both the strong and the weak interactions, and neutron evaporation from the nuggets might be significant to make also a neutron-rich environment for the nucleosynthesis of heavy nuclei.
Because of a high $M_\rm{max}\sim 3M_\odot$, the merger remnant of
binary strangeon stars would usually be very long-lived (even to be
stable if $M<M_\rm{max}$, rather than collapsing quickly to a black
hole), that would additionally power both the fire-balls of GRB
(gamma-ray burst) and KN during cooling and spinning
down~\cite{DL98,Lai18,Piro18,Xue19}, though the details of free
energy and its release are still a matter of debate.

Third, the Ruderman-Sutherland (i.e., RS75~\cite{RS75}) model is
still popular to connect radio magnetospheric activity with general
observations, having a ``user friendly'' nature, given that the pulsar
radiative mechanism is not well understood even though after more half
a century of observations. Nonetheless, if pulsars are strangeon stars, both the
binding-energy problem (for antiparallel rotator, $\bf{\Omega\cdot
\mu}<0$, with $\Omega$ the angular velocity of rotation and $\mu$ the magnetic dipolar momentum) and the antipulsar embarrassment (for parallel rotator,
$\bf{\Omega\cdot \mu}>0$) would not exist anymore~\cite{Xu98,XQZ99}, and then
RS75 works well.
It is worth noting that the sparking dynamics depends on the
surface roughness of pulsar, and the correlation ($P_2-P_3$) of subpulse-drifting pulsar PSR B2016+28 could be evidence for strangeon star~\cite{LuJG19}.
Extremely relativistic electron/positron particles collide the pulsar surface, with kinematic energy of $\sim \gamma m_{\rm e}c^2\sim 1$ TeV (Lorentz factor $\gamma\sim 10^6$), and the energy in the center of mass for two electron/positron collision is even $\sqrt{2\gamma}m_{\rm e}c^2\sim 1$ GeV.
What if atom matter of normal neutron star reacts in such an accelerator? High-energy reactions are too utmost to keep a stable electric matter surface even with extremely high magnetic field, but a strangeon surface with ``small mountains'' could stand against the bombardments.
Anyway, more theoretical researches on these topics (Lorentz factor
much higher than $\gamma\sim 10$, with which an investigation has
been done~\cite{BPO19}) are interesting and necessary, but most
importantly, an elaborated observation to trace the sparking points
(a small mountain on rough cap has priority to discharge) on polar
gap is surely welcome.

\vspace{2mm}
\noindent%
2. {\em Strangeon Cosmic Ray}.
In the regime of free quarks, Witten~\cite{Witten1984} conjectured
an absolutely stable state of strange {\em quark} matter, and
addressed dramatic consequences of this strong matter: quark star
produced during supernova, quark nuggets residual after cosmic QCD
phase-transition, as well as strange cosmic ray.
These three are retained if quarks are not free but localized in
strangeons, and strangeon matter shares a similar trinity: compact
star, dark matter, and cosmic ray (besides the strangeon star/planet
issues discussed above).

Stable strangeon nuggets with baryon number $A\gtrsim A_{\rm c}$
could be ejected relativistically/non-relativistically after a merge
of binary strangeon stars. While nuggets with $A<A_{\rm c}$ may
decay quickly into {\bf 2}-flavoured nucleon matter and power the KN
radiation, those with $A>A_{\rm c}$ may fly away and reach the Earth
through long-time travel, eventually resulting in a strangeon cosmic
ray air-shower in Earth's atmosphere.
For instance, the rest mass of a nugget with $A\sim 10^{10}$ is
$\sim 10^{19}$ eV, and the deposit energy during corresponding air
shower could then be of order $10^{18\sim 20}$ eV, depending
certainly on its speed.
This kind of strangeon cosmic ray remains a possibility up-to-now,
for the existing experiments are only sensitive to strange nuggets
with $A<10^5$.

Let's briefly discuss the air-shower of strangeon cosmic ray.
For Lorentz factor $\gamma<2$, the kinematic energy of strangeon cosmic ray is $E_\rm{CR}\sim m_\rm{CR}c^2\beta^2$, where $\beta=\sqrt{1-\gamma^{-2}}$ measures the speed and $m_\rm{CR}$ is the rest mass. One can then have
\begin{equation}
E_\rm{CR}\sim (10^{17}\rm{eV})~A_{10}\beta_{0.1}^2,
\label{E_CR}
\end{equation}
where the baryon number $A=A_{10}\times 10^{10}$ and $\beta=0.1\beta_{0.1}$.
However, in the cosmic ray rest frame, a proton in Earth's atmosphere has a kinematic energy of
\begin{equation}
E_\rm{proton}\sim (10\rm{MeV})~\beta_{0.1}^2.
\label{E_proton}
\end{equation}
A hadronic cascade may stop when $E_\rm{proton}<m_\pi c^2\sim
100$MeV (or $\beta<0.3$), and an electromagnetic cascade may end
when $E_\rm{proton}<2m_{\rm e} c^2\sim 1$MeV (or $\beta<0.03$).
Thus, the interaction should become very weak when the speed of
strangeon nugget is lower than $\sim 10^9$ cm/s, just going almost
freely through the Earth.
In case of $\beta>0.1$, an atomic nucleus would be destroyed during collision, for the nuclear binding energy per baryon is comparable to $E_\rm{proton}$.
Assuming deposit PeV-energy in air-shower and $\sim 100$ MeV-energy lose per nucleon during interaction, we may estimate the atmospheric depth, $X$, by $10^8\rm{eV}\cdot (X/m_{\rm p})\cdot (A^{1/3}\rm{fm})^2\sim 10^{15}\rm{eV}$,
\begin{equation}
X\sim 10^7 A^{-2/3} m_{\rm p}\sim (400~\rm{g/cm}^2)\cdot
A_{10}^{-2/3}.%
\label{X}
\end{equation}
Without a doubt, it is worth waiting for an identification of strangeon cosmic ray either in low altitude (e.g., the Pierre Auger Observatory) or in high altitude (e.g., the Large High Altitude Air Shower Observatory, abbreviated to LHAASO).

\vspace{2mm}
\noindent%
3. {\em Strangeon Dark Matter}.
The quark coupling in strangeon matter is stronger than in strange
quark matter, and the cosmic QCD phase transition could then
be of first order. Therefore, the resultant quark nuggets would thus
survive after boiling and evaporation, manifested in the form of
invisible ``dark matter''.

It is still a challenge to know the nature of dark matter even after
more than a century.
It is a general view point that dark matter represents a glimpse of
``new'' physics beyond the standard model, but strangeon
dark matter does exist in the standard model of particle physics~\cite{LX10}.
In contrast to WIMPs (weakly interacting massive particle, such as
supersymmetric particle and axion), strangeon dark matter could be attractive to understand
the comparable ratio of dark matter to normal baryonic matter, $\sim 5:1$ (only order of one).
Additionally, non-relativistic strangeon nugget with $A\sim 10^{30}$
is certainly of cold dark matter candidates, but conventional
experiments to directly detect dark matter are {\em not} sensitive
to strangeon dark matter.
While today's experiments provide serious challenge to some of the
popular dark matter models, it will become more challenging when the
sensitivity of future experiments reaches the so-called neutrino
floor.
Therefore, strangeon dark matter could be paid attention to as a new
insight into dark matter if current detectors with increasing
sensitivity would further fail to catch the elusive dark matter
particle.

We may solve the lithium-problem with strangeon dark matter.
Big bang nucleosynthesis (BBN) theory predicts the $^7$Li abundance,
which is about 3 times larger (around $5-\sigma$ mismatch) than that observed, and it is
thought that the destruction of $^7$Be could be a promising clue
(e.g., $^7$Be($\alpha$, $\gamma$)$^{11}$C~\cite{Li18}). Note that
$^7$Be decays with a half-life of 53.22 days via $^7$Be + e
$\rightarrow ^7$Li + $\nu_{\rm e}$, and the $^7$Be abundance was
about ten times of the $^7$Li abundance during the early age of $\lesssim 1$
day $\sim 10^5$ s.
Among the 12 main BBN reactions, only 4 of them could be responsible for the destruction of nuclear species when they collide strangeon nugget.
As shown in Table 1, the binding of $^7$Be is relatively weak
during the BBN stage in which the kinematic energy is order of 0.1 MeV, at which the
deuteron photon-disintegration ceases (i.e., deuterons frozen-out), so that a $^7$Be-nucleus could fragmentate into $^4$He
and $^3$He when colliding with a strangeon nugget (note: the binding
energy of the first reaction is smaller than the second in Table 1),
but the abundances of other nuclear species may change
insignificantly.
Both a hot environment (i.e., energetic photons) and the collision energy in the center of mass (being order of 0.1 MeV because of massive strangeon nugget) would affect the fragmentation process.
Certainly a quantitative calculation of this destruction of $^7$Be would be interesting.
\begin{table}[tbh]
\caption{Binding Energy of a few nuclear reaction.}
{\normalsize\begin{tabular}{lrr}
\hline \hline Reaction & Product & Binding Energy/MeV\\
\hline %
$^4$He~+~$^3$He & $^7$Be & 1.58713\\
$^1$H~+~n & D & 2.22457\\
$^4$He~+~$^3$H & $^7$Li & 2.30123\\
$^1$H~+~$^2$H & $^3$He & 5.49348\\
\hline \hline
\end{tabular}}
\end{table}

Besides, strangeon dark matter is naturally self-interaction, and it
could be a useful way for us to understand the observations
that dark matter halos of some dwarf galaxies are less dense in
their central regions compared to expectations from collisionless
N-body simulations.
By fitting the rotation curves of a sample of galaxies in a self-interacting dark matter model, authors found a velocity-dependent value of $\sigma/m\sim (0.1 \sim 3)$~cm$^2$/g
~\cite{KTY16,KKPY17}, where $\sigma$ is the scattering cross section and $m$ is the dark matter particle mass.
For a charged strangeon nugget with baryon number $A=A_{30} 10^{30}$, its mass is $m\simeq 1.6\times 10^6A_{30}$ g, and the Debye length, at the scale of which the nugget's electric fields are screened in interstellar medium, is $\lambda_{\rm D}=\sqrt{kT/(4\pi n e^2)}$, with $T$ the temperature and $n$ the number density of free particles charged.
We can then estimate $\sigma/m$ for self-interacting strangeon dark matter,
\begin{equation}
(\sigma/m)_\rm{strangeon dark matter}\simeq \lambda_{\rm D}^2/m \sim (3~\rm{cm}^2/{\rm g})\cdot T_5 n_1 A_{30}^{-1},
\end{equation}
where $T=T_5\times 10^5$ K and $n=n_1\times 1$~cm$^{-3}$. This value estimated would be comparable to the observations.

How can we detect directly strangeon dark matter?
A series of weak moon-quakes could occur when a strangeon nugget penetrates the Moon, with a rate of $\sim 6/A_{30}$ yr$^{-1}$ for strangeon dark matter with dynamical velocity of $\sim 200$ km/s near the Sun~\cite{Xu18}.
An observatory on the quiet moon to monitor its weak quakes would help.

\section{Conclusions}

It is a great achievement to recognize microscopically that all
objects are composed by ``uncuttable'' atoms during the ancient
Greece time of Democritus, but an atom is actually consist of a
nucleus within an electron cloud.
Nucleons are the deeper composition of a nucleus, but they consist
of only two flavours of valence quarks though there are totally six
flavours of quarks in the standard model of particle physics.
Would it be possible to build a kind of stable condensed matter with
nucleon-like units being 3-flavored?
This is the story focused in this contribution, saying that
3-flavored strangeons might constitute macroscopic even cosmic
strong matter, and that strangeon matter could be manifested as
trinity: strangeon star/planet, strangeon cosmic ray and strangeon
dark matter.

The inner structure of pulsar-like compact objects as well as the
EoS of supranuclear dense matter are challenging in both physics and
astronomy, and this is really the motivation that we have the Xiamen
EoS Workshop in the new era of gravitational wave astronomy.
We think that both perturbative and non-perturbative QCD effects are
responsible to solving the problem: the former results in
quark-flavour maximization, and the latter contributes to the
localization of quarks in strong unit (nucleon or strangeon) as well as to a hard core
between the units.
One may speculate a state of strange quark matter if only the former
is considered (i.e., the Witten conjecture), but the latter does
play a key role in determining the real state. Nevertheless, a
strangeon matter conjecture comes now, with the inclusion of both
kinds of the QCD effects.
Among the compact star models in the academic market, listed in
Table 2, neutron stars are {\bf 2}-flavoured asymmetric but strange
stars (including strange quark star and strangeon star) are {\bf
3}-flavoured symmetric (i.e., strange stars are more symmetrical in quark flavours than neutron star).
However, it is still questionable whether quarks could be free at an energy scale of $E_{\rm{scale}}\sim 0.5 \rm{GeV}<\Lambda_\chi$.
If quarks are localized in strangeon, and additionally a hard core
exists between strangeons due to rich non-perturbative QCD effects,
then the EoS should be very stiff and, due to {\bf 3}-flavour
maximization, the energy per baryon could also be the lowest.
\begin{figure}[h!]
\centering
  \includegraphics[width=12cm]{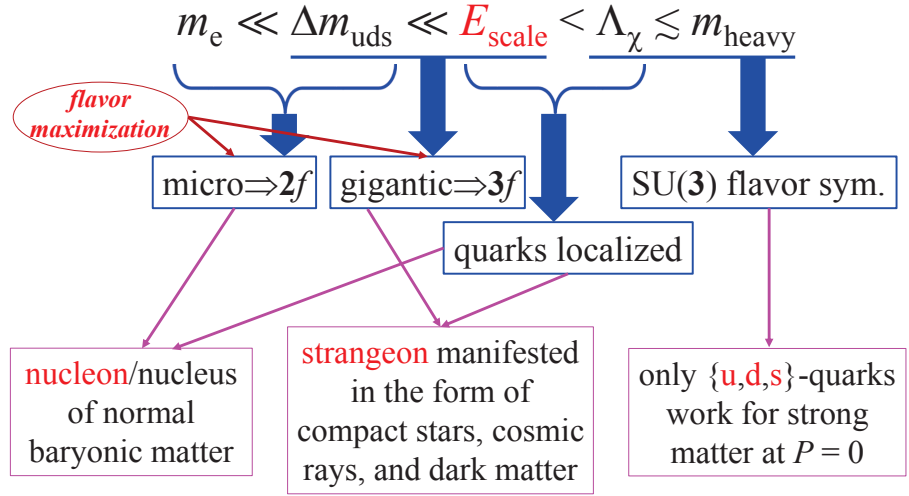}
  \caption{%
  The states of matter in the Universe depend on different energy scales.
  We are lucky to have {\bf 2}-flavoured atoms because the electromagnetic force is much
  weaker than the strong force, so that electron clouds are always much bigger than the Compton wavelength, $h/(m_{\rm e}c)=0.024\AA$, and that the electron's non-relativistic role is not energetically important if $s$-quark decays into $u/d$-quark via the weak interaction for micro-strong matter, the nuclei. However, electrons become extremely energetic when a huge number of nuclei are squeezed into close contact, and {\bf 3}-flavour-symmetry restoration may occur for this macro-strong matter (baryon number $A>10^9$), resulting in a conversion of the basic unit from {\it nucleon} to {\it strangeon} by the weak interaction.
  }%
  \label{scale}
\end{figure}

\begin{table}[tbh]
\caption{Compact star models: a comparison.}
{\normalsize\begin{tabular}{lr||lr|r|r}
\hline\hline Models& Basic unit & Flavour & Asymmetry & Quark coupling, EoS&Surface binding\\
\hline %
Neutron Star & nucleon & {\bf 2} ($u$ \& $d$) & $\delta>0.8$&strong, stiff if no hyperon&gravity\\
Strange Quark Star&quark&{\bf 3} ($u$, $d$ \& $s$)&$\delta< 10^{-4}$&weak, softened with $s$&self strong force\\
Strangeon Star &strangeon&{\bf 3} ($u$, $d$ \& $s$)&$\delta< 10^{-4}$&strong, stiff in any case&self strong force\\
\hline\hline
\end{tabular}}
\end{table}
The existence of the present universe may depend on a few fundamental parameters, as summarized in Fig.~\ref{scale}.
Heavy flavours of quarks will not participate in pressure-free strong matter where the coupling between quarks is so strong that quarks are localized either in nucleon ({\bf 2}-flavoured) or strangeon ({\bf 3}-flavoured).
Micro-strong matter could be only {\bf 2}-flavoured as the weak
interaction can convert $s$-quark to $u/d$-quark, even that Nature
loves a principle of flavour maximization, but macro-strong matter
should be {\bf 3}-flavoured otherwise the system would be unstable
due to a high energy of electron or a high nuclear symmetry energy.

\section{Acknowledgement}
The author would like to thank those involved in the continuous discussions in the pulsar group at Peking University. This work is supported by the National Key R\&D Program of China (No. 2017YFA0402602), the National Natural Science Foundation of China (Grant Nos. 11673002 and U1531243), and the Strategic Priority Research Program of CAS (No. XDB23010200).



\begin{thebibliography}{1}

\bibitem{Xu03}
R. X. Xu, ApJ. \textbf{596}, L59 (2003).

\bibitem{Xu18}
R. X. Xu, Sci. China-Phys. Mech. Astron. \textbf{61}, 109531 (2018).

\bibitem{Landau32}
L. Landau, Sov. Phys. \textbf{1}, 285 (1932).

\bibitem{Bodmer1971}
A. Bodmer, Phys. Rev. D \textbf{4}, 16 (1971).

\bibitem{Witten1984}
E. Witten, Phys. Rev. D \textbf{30}, 272 (1984).

\bibitem{xg2017}
R. X. Xu, and Y. J. Guo, 2017, in: ``centennial of general relativity - a celebration'', Ed. Cesar A. Zen Vasconcellos, World Scientific Publishing Company, p.119-146 (arXiv:1601.05607).

\bibitem{Wang2017}
W. Y. Wang, J. G. Lu, H. Tong, et al., ApJ. \textbf{837} 81 (2017).

\bibitem{Xu2015}
R. X. Xu, Acta Astron. Sinica. \textbf{56} {\em Suppl.}, 82 (2015) (arXiv:1507.07172).

\bibitem{Tabor91}
D. Tabor, Gases, liquids and solids -- and other state of matter,
Cambridge Univ. Press, Cambridge, 1991.

\bibitem{Wilczeck2007}
F. Wilczeck, Nature. \textbf{445}, 156 (2007).

\bibitem{LX09}
Y. Lai, R. X. Xu, Mon. Not. R. Astron. Soc.-Lett. \textbf{398}, L31
(2009).

\bibitem{LaiXu17}
X. Lai, R. Xu, J. of Phys: Conference Series. \textbf{861}, 012027
(2017).

\bibitem{LuXu17}
J. Lu, R. Xu, JPS Conf. Proc. \textbf{20}, 011026 (2017).

\bibitem{tc99}
S. E. Thorsett, D. Chakrabarty, ApJ. \textbf{512}, 288 (1999).

\bibitem{Tauris11}
T. M. Tauris, N. Langer, M. Kramer, MNRAS. \textbf{416}, 2130 (2011).

\bibitem{0740}
H. T. Cromartie, E. Fonseca, S. M. Ransom, et al., Nature Astronomy (submitted,
arXiv:1904.06759).

\bibitem{XW03}
R. X. Xu, F. Wu, Chin. Phys. Lett. \textbf{20}, 806 (2003).

\bibitem{Horvath12}
J. E. Horvath, Res. Astron. Astrophys. \textbf{12}, 813 (2012).

\bibitem{Xu14}
R. X. Xu, Res. Astron. Astrophys. \textbf{14}, 617 (2014).

\bibitem{Lai18}
X. Y. Lai, Y. W. Yu, E. P. Zhou, et al., RAA. \textbf{18}, 24 (2018).

\bibitem{LZX19}
X. Y. Lai, E. P. Zhou, R. X. Xu, EPJA. accepted (arXiv:1811.00193).

\bibitem{DL98}
Z. G. Dai, T. Lu, Phys. Rev Lett. \textbf{81}, 4301 (1998).

\bibitem{Piro18}
L. Piro, E. Troja, B. Zhang, et al., MNRAS. \textbf{483}, 1912
(2018).

\bibitem{Xue19}
Y. Q. Xue, X. C. Zheng, Y. Li, et al., Nature. \textbf{568}, 198 (2019).

\bibitem{RS75}
M. A. Ruderman, P. G. Sutherland, ApJ. \textbf{196}, 51 (1975).

\bibitem{Xu98}
R. X. Xu, G. J. Qiao, Chin. Phys. Lett. \textbf{15}, 934 (1998).

\bibitem{XQZ99}
R. X. Xu, G. J. Qiao, B. Zhang, ApJ. \textbf{522}, L109 (1999).

\bibitem{LuJG19}
J.-G. Lu, B. Peng, R.-X. Xu, et al., Sci. China-Phys. Mech. Astron. \textbf{62}, 959505 (2019).

\bibitem{BPO19}
M. Baun\"ock, D. Psaltts, F. \"Ozel, ApJ. \textbf{872}, 162 (2019).

\bibitem{LX10}
X. Y. Lai, R. X. Xu, J. Cosmol. Astropart. Phys. \textbf{5}, 028
(2010, arXiv: 0911.4777).

\bibitem{Li18}
M. Hartos, C. A. Bertulani, Shubhchintak, et al., ApJ. \textbf{862}, 62 (2018)

\bibitem{KTY16}
M. Kaplinghat, S. Tulin, H.-B. Yu, Phys. Rev. Lett. \textbf{116},
041302 (2016).

\bibitem{KKPY17}
A. Kamada, M. Kaplinghat, A. B. Pace, H.-B. Yu, Phys. Rev. Lett.
\textbf{119}, 111102 (2017).



\end{thebibliography}

\end{document}